\documentclass[aps,twocolumn,showpacs,preprintnumbers,showkeys]{revtex4}

\usepackage[english]{babel}
\usepackage{dcolumn}
\usepackage{latexsym}
\usepackage{amsmath}
\usepackage{amssymb}
\usepackage{graphics}
\usepackage{float}
\usepackage{placeins}
\usepackage[active]{srcltx}

\begin{document}

\title{Molecular structures in charmonium spectrum:
The $XYZ$ puzzle}
\author{P. G. Ortega, D. R. Entem and F. Fern\'andez}
\affiliation{Grupo de F\'isica Nuclear and IUFFyM,
Universidad de Salamanca, E-37008 Salamanca, Spain}

\begin{abstract}
We study in the framework of a constituent quark model the possible contributions of molecular 
structures to the $XYZ$ charmonium like states. We analyze simultaneously the 
$c\bar{c}$ structures and the possible molecular components in a formalism which 
allows us to treat channels below and above thresholds. The only molecular 
state found in the $1^{++}$ sector correspond to the $X(3872)$. Molecular 
resonances also appear with other quantum numbers. So, the so called $Y(3940)$ and 
the $X(3915)$ are suggested to be $J^{PC}=0^{++}$ charmonium states. In the 
$J^{PC}=1^{--}$ sector we also found significant contributions of the molecular 
structures which can affect the phenomenology.
\end{abstract}


\maketitle

\section{Introduction.}

In the last few years several charmonium like states were observed with similar 
masses near $3.9\,GeV$ but with quite different properties and in very different 
production processes. Altogether these states were called the $XYZ$ states.
A complete list of these new states can be found in Ref.~\cite{nora}. Among them
we will only comment on the confirmed states.

This new charmonium era started around 2003 when the Belle Collaboration discovered
the lightest one, the  $X(3872)$, in the exclusive decay $B^{\pm}\to 
K^{\pm}\pi^{+}\pi^{-}J/\psi$~\cite{Choi03}. The mass of the state was measured 
to be $3872.0\pm0.6$ MeV very close to the $M_{D^{0}}+M_{D^{\ast0}}$ 
threshold. The width was found to be very small $\Gamma<2.3$ MeV. The state 
was soon confirmed by CDF~\cite{Ac04}, D0~\cite{Ab04} and BaBar~\cite{Au04}. 
By combination of  
the recent results reported by the
Belle~\cite{BelleDJPsi}, BaBar~\cite{BaBarDJPsi} and CDF~\cite{CDFDJPsi} 
Collaborations,
the mass value is established at $M_X=3871.55\pm 0.20$ MeV. 

An striking feature of the $X(3872)$, which cannot be explained by a simple 
$c\bar c$ structure, is the ratio~\cite{m9}
\begin{eqnarray}
R_1= \frac{X(3872)\rightarrow \pi^+\pi^-\pi^0J/\psi}{X(3872)\rightarrow 
\pi^+\pi^-J/\psi}&=&0.8\pm 0.3.
\end{eqnarray}
The dipion mass spectrum in the
$\pi^+ \pi^- J/\psi$ channel shows that the pions come from the $\rho^0$
resonance. On the other hand the $\pi^+\pi^-\pi^0$ mass spectrum 
has a strong peak around
$750$ MeV suggesting that the process is dominated by an $\omega$
meson. 
Although this number should be corrected by the strong phase suppression of the 
$\omega J/\psi$ channel against the  $\rho J/\psi$ one, the ratio $R_1\sim1$ is 
incompatible with a traditional charmonium assumption and is telling us that 
some isospin mixing is needed and that this mixing requires the contribution of both 
neutral and charged $DD^*$ channels.
Recently the Belle Collaboration~\cite{XL5} measured the ratio
\begin{equation} 
R_2=\frac{\Gamma(X(3872)\to\gamma\Psi(2S))}{\Gamma(X(3872)\to\gamma 
J/\psi)}\leq 2.1 (\mbox{at 90}\% \mbox{ C.L.}) 
\end{equation}
which complicates the interpretation of this state.

Two years later, the so called at that time $Y(3940)$ was observed by the Belle 
Collaboration as a near-threshold enhancement in the $\omega J/\psi$ invariant 
mass distribution for the $B\rightarrow K\omega J/\psi$
decay~\cite{Choi05}. Belle reported a mass of $M=3943\pm 11\pm 13$ MeV and a width 
$\Gamma =87\pm 22\pm 26$ MeV. Belle observation seems to be confirmed by BaBar~\cite{Au08}, 
although the mass ($M=3914\pm 4.1$ MeV) and the width ($\Gamma 
=29\pm 10$ MeV) were both smaller than Belle values. A later measurement of 
BaBar~\cite{Amo10} confirmed this last mass value ($M=3919\pm 3.8\pm 2.0$ MeV).
Very recently a new charmonium-like state, the $X(3915)$, has been reported by 
the Belle Collaboration in the $\gamma\gamma\to J/\psi\omega$ decay~\cite{ueh10}. 
The measured mass is $M=3914\pm3\pm2$ MeV and the width $\Gamma=23\pm9$ MeV. 
It has not yet been seen in the $DD$ channel. Despite of the different 
mass and width of the first measurement of Belle, some authors~\cite{nora,nak10} 
subsumed both states under the name $X(3915)$,
although the question if there is one or two different resonances is not definitely settled.

Another charmonium like state, the $X(3940)$, was observed in this region by 
Belle as a resonance in the double charmonium production $e^{+}e^{-}\to J/\psi 
D{D}^{\ast}$ in the mass spectrum recoil against the $J/\psi$~\cite{Ab07}. 
Later on Belle confirmed the observation of $X(3940)\to D{D}^{\ast}$ 
decay~\cite{Pak08}. In addition Belle found a new charmonium like state $X(4160)$ 
decaying into $D^*D^*$. Neither of them have been seen in the experimentally more 
accessible $DD$ channel.
As the decay of the $Y(3940)\to D{D}^{\ast}$ was not observed in the $B\to 
Y(3940)K$~\cite{Ad08}, the $X(3940)$ and the $Y(3940)$ should be different 
states.

Two more states increase the experimental findings in this region. The 
$Z(3930)$ was reported by Belle in the $DD$ channel produced in 
$\gamma\gamma$ collisions with mass and width $M=3929\pm6$ MeV and 
$\Gamma=29\pm10$ MeV~\cite{ueh06}. 

Finally, it is worth to mention that one more resonance has been classified as 
'well established' in Ref.~\cite{nora}. It was found by the BaBar Collaboration~\cite{XL1} 
in the reaction $e^+e^-\rightarrow D D$ with a mass of $3943\pm 
17\pm 12$ MeV and a width of $52\pm 8\pm 7$ MeV and confirmed by Belle~\cite{XL2}. 
It was called $G(3900)$ in Ref.~\cite{XL1}.

Concerning the quantum numbers of the new states, measurement of the angular 
correlations between final state particles in the $X(3872)\to 
\pi^{+}\pi^{-}J/\psi$ decay~\cite{abe05} together with small phase space 
available for the decay $X(3872)\to D^{0}\bar{D}^{0}\pi^{0}$~\cite{gokhroo06} 
strongly favors the $J^{PC}=1^{++}$ quantum numbers for this state. However 
some recent results seems to favor negative parity for this meson~\cite{Amo10}.

The situation is worse in the case of the $X(3940)$. It has not been seen in 
the $DD$ channel which rules out the $J^{P}=0^{++}$ assignment. The 
dominant $DD^{\ast}$ mode suggest that the $X(3940)$ is the 
$c\bar c(2^{3}P_{1})$ state with $J^{PC}=1^{++}$ but this quantum numbers 
seems to be assigned to the $X(3872)$. 

More consensus exists with the assignment of the $Z(3930)$. The $DD$ decay 
mode makes it impossible to be the $\eta_{c}(3S)$ $J^{PC}=0^{-+}$ state. The 
two photon production can only produce $DD$ states in $0^{++}$ or 
$2^{++}$ and these two cases can be distinguished looking to the 
$dN/dcos\theta$ distribution being $\theta$ the angle between the incoming 
photon and the $D$ meson in the $\gamma\gamma$ center of mass system. This 
distribution is flat for $0^{++}$ states and behaves like $sin^{4}\theta$ for 
$2^{++}$. The measurement of Belle~\cite{ueh06} strongly favors the $2^{++}$ 
case.

The $J^{PC}$ assignment in the case that the $Y(3940)$ was a $c\bar{c}$ state is 
still unclear. A conventional $c\bar{c}$ charmonium interpretation is in 
principle disfavored since it is well above the threshold for open charm decays 
and then these decay modes would dominate over the $\phi J/\psi$ and 
$\omega J/\psi$ decay rates.
The $X(3915)$ is preliminary assigned to be $0^{++}$ or $2^{++}$ ~\cite{ueh10}.

Finally due to the entrance channel of the production reaction the $G(3900)$ is 
clearly a $J^{PC}=1^{--}$ state.

Obviously it is difficult to accommodate all these states in a $q\bar q$ scheme 
and all type of hypothesis about their structures (molecules, hybrids, 
tetraquarks) has been proposed in the literature (see ~\cite{Dre10} for a 
review). 

The ${P=+}$, ${C=+}$ sector is specially suited for the coexistence of $c\bar{c}$ 
states and molecular structures. The reason is the following: taken into account 
the negative intrinsic parity of the quark-antiquark pair, to get a positive 
parity state one needs at least one unit of angular momentum. However four 
quarks can reach the same positive parity with zero angular momentum. Then, the 
energy increase due to the angular momentum excitation may compensate the two 
additional light quark masses making the $c\bar{c}$ and the $c\bar{c}q\bar{q}$ 
structures almost degenerate. 

This mechanism has been suggested in~\cite{ortega08} as a possible explanation 
to the $X(3872)$  properties. In this reference a coupled channel calculation 
of the $1^{++}$ sector has been performed including $c\bar{c}$ and  $DD^*$ 
molecular configurations. The $X(3872)$ appears as a dynamically generated 
$c\bar{c}$ and  $DD^*$ molecule by the coupling to a $\chi_{c_1}(2P)$ quark 
state. Although the $c\bar{c}$ mixture is less than $10\%$, it is important 
to bind the molecule. In addition  $\pi^+\pi^- J/\psi$ decay modes data from 
Belle and BaBar are reasonably explained.

In this paper we will propose a theoretical explanation of the nature of some 
of these states using the same constituent quark model of Ref.~\cite{ortega08}. 
This has been successfully used to describe  hadronic spectroscopy and hadronic 
reactions~\cite{h1,h2,h3,h4} and is based on the assumption that constituent 
quark mass is a consequence of the spontaneous chiral symmetry breaking and  
has been recently applied to the study of the energy spectrum and decay 
properties of the $J^{PC}=1^{--}$ charmonium sector~\cite{Seg08}. Taken into 
account the existence of several thresholds in this energy region we develop a 
formalism which treats simultaneously molecular states above and below 
the different thresholds. This allows us to consider both dynamically generated 
molecular states by the coupling with $c\bar{c}$ structures and molecular 
components in the dressed $c\bar{c}$ states.
 
The paper is organized as follow. In Section~\ref{model} we first discuss the basic 
ingredients of the constituent model, the coupled channel formalism and the 
coupling mechanism between the different channels. Results and comments are 
given in Section~\ref{results}. Finally we summarize the main achievements of our 
calculation in Section~\ref{summary}.

\section{The Model}
\label{model}

\subsection{The constituent quark model}

The constituent quark model used in this work has been extensively
described elsewhere~\cite{h3} and therefore 
we will only summarize here its most
relevant aspects. The chiral symmetry of the original QCD Lagrangian appears
spontaneously broken in nature and, 
as a consequence, light quarks acquire a
dynamical mass. The simplest Lagrangian invariant under chiral rotations
must therefore contain chiral fields, and can be expressed as 
\begin{equation}
\label{lagrangian}
{\mathcal L}
=\overline{\psi }(i\, {\slash\!\!\! \partial} -M(q^{2})U^{\gamma_{5}})\,\psi 
\end{equation}
where  $U^{\gamma_5}=e^{i\frac{
\lambda _{a}}{f_{\pi }}\phi ^{a}\gamma _{5}}$ is 
the Goldstone boson fields matrix and $M(q^2)$ the dynamical
(constituent) mass. 
This Lagrangian has been derived in Ref.~\cite{r17}
as the low-energy limit in the instanton liquid model. In this model
the dynamical mass vanishes at large momenta and it is frozen at low 
momenta, for a value around 300 MeV. 
Similar results have also been obtained in lattice
calculations~\cite{lat}. 
To simulate this behavior we parametrize 
the dynamical mass as
$M(q^{2})=m_{q}F(q^{2})$, 
where $m_{q}\simeq $ 300 MeV, and
\begin{equation}
F(q^{2})=\left[ \frac{\Lambda _{\chi}^{2}}{\Lambda _{\chi}^{2}+q^{2}}%
\right] ^{
{\frac12} 
} \, .
\end{equation}
The cut-off $\Lambda _{\chi}$ fixes the
chiral symmetry breaking scale. 

The Goldstone boson field matrix $U^{\gamma_{5}}$ can be expanded in terms of 
boson fields,
\begin{equation}
U^{\gamma_{5}}=1+\frac{i}{f_{\pi }}\gamma^{5}\lambda^{a}\pi^{a}-\frac{1}{%
2f_{\pi}^{2}}\pi^{a}\pi^{a}+...
\end{equation}
The first term of the expansion generates the constituent quark mass while the
second gives rise to a one-boson exchange interaction between quarks. The
main contribution of the third term comes from the two-pion exchange which
has been simulated by means of a scalar exchange potential.

In the heavy quark sector 
chiral symmetry is explicitly broken and this type of interaction does not act. 
However it constrains the model parameters through the light meson 
phenomenology 
and provides a natural way to incorporate the pion exchange interaction in the 
open charm dynamics.

Below the chiral symmetry breaking scale quarks still interact
through gluon exchanges  described by
the Lagrangian
\begin{equation}
\label{Lg}
{\mathcal L}_{gqq}=
i\sqrt{4\pi \alpha _{s} }\,\,\overline{\psi }\gamma _{\mu }G^{\mu
}_c \lambda _{c}\psi  \, ,
\end{equation}
where $\lambda_{c}$ are the SU(3) color generators and $G^{\mu}_c$ the
gluon field. 
The other QCD nonperturbative effect corresponds to confinement,
which prevents from having colored hadrons.
Such a term can be physically interpreted in a picture in which
the quark and the antiquark are linked by a one-dimensional color flux-tube.
The spontaneous creation of light-quark pairs may
give rise at same scale to a breakup of the color flux-tube~\cite{Bali}. This 
can be translated
into a screened potential~\cite{c28} in such a way that the potential
saturates at the same interquark distance
\begin{equation}
V_{CON}(\vec{r}_{ij})=\{-a_{c}\,(1-e^{-\mu_c\,r_{ij}})+ \Delta\}(\vec{%
\lambda^c}_{i}\cdot \vec{ \lambda^c}_{j}).
\end{equation}
Explicit 
expressions for these interactions are given in Ref.~\cite{r16}.

\subsection{The coupled channel calculation}

In this section we present the formalism for the coupling of molecular 
structures with the $c\bar c$ spectrum.
We start defining the meson wave functions we will use all along the paper.
To found the quark-antiquark bound states we solve the 
Schr\"odinger equation using the Gaussian Expansion Method~\cite{r20}. 
In this method the radial wave functions solution of the Schr\"odinger equation 
are expanded in terms of basis functions
\begin{equation}
R_{\alpha}(r)=\sum_{n=1}^{n_{max}} b_{n}^\alpha \phi^G_{nl}(r)
\end{equation} 
where $\alpha$ refers to the channel quantum numbers.
The coefficients $b_{n}^\alpha$ and the eigenenergy $E$ are determined from the 
Rayleigh-Ritz variational principle
\begin{equation}
\label{RGM}
\sum_{n=1}^{n_{max}} \left[\left(T_{n'n}^{\alpha'}-EN_{n'n}^{\alpha'}\right)
b_{n}^{\alpha'}+\sum_{\alpha}
\ V_{n'n}^{\alpha'\alpha}b_{n}^{\alpha}=0\right]
\end{equation}
where the operators $T_{n'n}^\alpha$ and $N_{n'n}^\alpha$ are diagonal and 
the only operator which mix the different channels is the potential 
$V_{n'n}^{\alpha\alpha'}$.
 
A crucial problem of the variational methods is how to choose the radial 
functions $\phi^G_{nl}(r)$ in order to have a minimal, but enough, number of 
basis functions. Following~\cite{r20} we employ gaussian trial functions whose 
ranges are in geometric progression. The geometric progression is useful in 
optimizing the ranges with a small number of free parameters. Moreover the 
distribution of the gaussian ranges in geometric progression is dense at small 
ranges, which is well suited for making the wave function correlate with short 
range potentials. The fast damping of the gaussian tail is not a real problem 
since we can choose the maximal range much longer than the hadronic size.

To model the $c\bar c$ system
we assume that the hadronic state is
\begin{equation} 
\label{ec:funonda}
 | \Psi \rangle = \sum_\alpha c_\alpha | \psi_\alpha \rangle
 + \sum_\beta \chi_\beta(P) |\phi_A \phi_B \beta \rangle
\end{equation}
where $|\psi_\alpha\rangle$ are $c\bar c$ eigenstates of the two body
Hamiltonian, 
$\phi_{M}$ are $q\bar q$  eigenstates describing 
the $A$ and $B$ mesons, 
$|\phi_A \phi_B \beta \rangle$ is the two meson state with $\beta$ quantum
numbers coupled to total $J^{PC}$ quantum numbers
and $\chi_\beta(P)$ is the relative wave 
function between the two mesons in the molecule. 
When we solve the four body problem we also use the gaussian expansion of the $q\bar q$
wave functions obtained from the solution of the two body problem. 
This procedure 
allows us to introduce in a variational way possible distortions of the two body 
wave function within the molecule. To derive the meson-meson interaction from 
the $qq$ interaction we use the Resonating Group Method (RGM). 

The coupling between the two sectors requires the creation of a light quark 
pair $n\bar n$. Similar to
the strong decay process this coupling should be in principle driven by the 
same interquark Hamiltonian which determines the spectrum. However Ackleh {\it et 
al.}~\cite{n17} have shown that the quark pair creation $^3P_0$ 
model~\cite{r21}, gives  similar results to the microscopic calculation. The 
model assumes that the pair creation Hamiltonian is 
\begin{equation} 
\mathcal{H}=g \int d^3x \,\, \bar \psi(x) \psi(x)
\end{equation}
which in the non-relativistic reduction is equivalent to the transition
operator~\cite{Bonnaz}
\begin{equation}
\begin{split}
\mathcal{T}=&-3\sqrt{2}\gamma'\sum_\mu \int d^3 p d^3p' \,\delta^{(3)}(p+p')\\
&\times \left[ \mathcal Y_1\left(\frac{p-p'}{2}\right) b_\mu^\dagger(p)
d_\nu^\dagger(p') \right]^{C=1,I=0,S=1,J=0}
\label{TBon}
\end{split}
\end{equation}
where $\mu$ ($\nu=\bar \mu$) are the quark (antiquark) quantum numbers and
$\gamma'=2^{5/2} \pi^{1/2}\gamma$ with $\gamma= \frac{g}{2m}$ is a 
dimensionless constant 
that gives the strength of 
the $q\bar q$ pair creation from the vacuum.
From this operator we define the transition
potential $h_{\beta \alpha}(P)$ within the $^3 P_0$ model as~\cite{Kala0} 
\begin{equation}
\label{Vab}
	\langle \phi_{M_1} \phi_{M_2} \beta | \mathcal{T}| \psi_\alpha \rangle =
	P \, h_{\beta \alpha}(P) \,\delta^{(3)}(\vec P_{\mbox{cm}})
\end{equation}
where $P$ is the relative momentum of the two meson state.

Adding the coupling with charmonium states we end-up with the coupled-channel 
equations
\begin{widetext}
\begin{eqnarray}\label{ec:Ec-Res}
c_\alpha M_\alpha +  \sum_\beta \int h_{\alpha\beta}(P) \chi_\beta(P)P^2 dP & 
= & E c_\alpha \nonumber \\
 \sum_{\beta}\int H_{\beta'\beta}(P',P)\chi_{\beta}(P) P^2 dP + 
 \sum_\alpha h_{\beta'\alpha}(P') c_\alpha & = & E \chi_{\beta'}(P')
\end{eqnarray}
\end{widetext}
where $M_\alpha$ are the masses of the bare $c\bar c$ mesons and
$H_{\beta'\beta}$ is the RGM Hamiltonian for the two meson
states obtained from the $q\bar q$ interaction.
Solving the coupling with the $c\bar c$ states we arrive to an Schr\"odinger-type
equation
\begin{widetext}
\begin{eqnarray}\label{ec:Ec1}
\sum_{\beta}\int 
\left(H_{\beta'\beta}(P',P)+V^{eff}_{\beta'\beta}(P',P)\right) 
\chi_{\beta}(P) {P}^2 dP = E \chi_{\beta'}(P') 
\end{eqnarray}
\end{widetext}
where
\begin{equation}
V^{eff}_{\beta'\beta}(P',P;E)=\sum_{\alpha}\frac{h_{\beta'\alpha}(P')
h_{\alpha\beta}(P)}{E-M_{\alpha}}.
\end {equation}

Our aim is to find molecular states  above and bellow thresholds in the same 
formalism. However,above the threshold we will find complex eigenenergies, 
where the imaginary part is related to the width of such states. In order to 
find the poles of the $T$ matrix we must be in the correct Riemann sheet, so we 
have to analytically continue all the potentials for complex momenta.
Once the analytical continuation is performed, the previous coupled channel 
equations can be solved through the $T(\vec p,\vec p',E)$ matrix, solution of 
the Lippmann-Schwinger equation,
\begin{widetext}
\begin{equation}\label{ec:tmat}
T^{\beta'\beta}(P',P;E)=V^{\beta'\beta}_T(P', P;E)+\sum_{\beta''}\int 
V^{\beta'\beta''}_T(P',P'';E)
\frac{1}{E-E_{\beta''}(P'')}T^{\beta''\beta}(P'',P;E)
\,P''^2 dP'' 
\end{equation}
\end{widetext}
where $V^{\beta'\beta}_T(P',P;E)=V^{\beta'\beta}(P',P)+V^{eff}_{\beta'\beta}(P',P;E)$, 
$V^{\beta'\beta}(P',P)$ is the RGM potential and 
$V^{eff}_{\beta'\beta}(P',P;E)$ is the effective potential due 
to the coupling to intermediate $c\bar c$ states. 

In this way we study the influence of the $c \bar c$ states on the dynamics
of the two meson states. This is a different point of view from the usually
found in the literature where the influence of two meson states (in general
without meson-meson interaction) in the mass and width of $c \bar c$ states
is studied~\cite{Kala0}. Our approach allows to generate new states through 
the meson-meson
interaction due to the coupling with $c\bar c$ states and to the underlying
$q\bar q$ interaction. 

The $T$ matrix of Eq.~(\ref{ec:tmat}) can be factorized as~\cite{Baru} 
\begin{widetext}
\begin{equation}
\label{ec:tmat1}
T^{\beta'\beta}(P',P;E)=T^{\beta'\beta}_V(P',P;E)
+\sum_{\alpha,\alpha'}\phi^{\beta'\alpha'}(P';E)
\Delta_{\alpha'\alpha}^{-1}(E)\phi^{\alpha\beta}(P;E) 
\end{equation}
\end{widetext}
with 
$\Delta_{\alpha'\alpha}^{-1}(E)=\left((E-M_{\alpha})\delta^{\alpha'\alpha}+\mathcal{
G}^{\alpha'\alpha}(E)\right)^{-1}$ being the propagator of the mixed state and 
$T^{\beta'\beta}_V(P',P;E)$ the $T$ matrix of the RGM potential excluding the 
coupling to the $c\bar c$ pairs.

The new functions $\phi^{\beta\alpha}(P;E)$ can be interpreted as the 
dressed $^3P_0$ vertex by the RGM meson-meson interaction and are defined as 
\begin{equation}
\phi^{\beta\alpha}(P;E)= h_{\beta\alpha}(P)-\sum_{\beta'}\int 
\frac{T^{\beta\beta'}_V(P,q;E)h_{\beta'\alpha}(q)}
{q^2/2\mu-E}\,q^2dq.
\end{equation}

Resonances will appear as poles of the $T$ matrix, namely as zeros of the 
inverse propagator of the mixed state. Therefore the resonance parameters are solutions 
of the equation 
\begin{equation}
	\label{ec:pole}
	\left|\Delta_{\alpha'\alpha}(\bar{E})\right|=\left|(\bar{E}-M_{\alpha})
	\delta^{\alpha'\alpha}+\mathcal{G}^{\alpha'\alpha}(\bar{E})\right|=0
\end{equation}
with $\bar{E}$ the pole position. This equation is solved by the Broyden 
method~\cite{brm}.

From the solution of (\ref{ec:pole}) we obtain the energy and the total width of the 
resonance. However, we are faced with the problem of the definition of the 
partial width. A similar problem arise when one try to define the mass and the 
width of an unstable particle in a gauge independent way~\cite{Grassi00}.
Let assume the case of a $c\bar c$ bound state coupled to two meson states. 
If the two mesons are below threshold we get a mass shift of the particle mass 
but if they are above threshold the mass is also renormalized by the coupling 
and now the particle becomes unstable and acquires a width. The conventional 
definition of mass and width are in this case.
\begin{equation} 
\label{ec:onshell}
\begin{array}{rcl}
 M&=&M_0-\Re(\mathcal{G}(M)),\\
\Gamma&=&2\dfrac{\Im(\mathcal{G}(M))}{1+\Re(\mathcal{G}(M))}
\end{array}
\end{equation}
where $M_0$ is the bare mass and $\mathcal{G}(E)$ is the two meson loop. The 
partial width is defined by decomposing the numerator of the width in 
Eq.~(\ref{ec:onshell}) into a sum of contributions of different two meson 
channels. However this cannot be done in our case because 
we use the complex value of the energy at the pole position
to define the mass and width of the state
\begin{equation} 
\bar E = M_0-\mathcal{G}(\bar E).
\end{equation}
With the usual parametrization $\bar E=M_r-i\Gamma_r/2$ we obtain
\begin{equation}\label{ww}
 \Gamma_r=2\Im(\mathcal{G}(\bar E))
\end{equation}
and we cannot follow the usual procedure to define the partial widths. Instead,
following Ref.~\cite{Grassi00}, we start from the $S$-matrix for an arbitrary 
number of $c\bar c$ states
\begin{widetext}
\begin{equation}
 S^{\beta'\beta}(E)=
S^{\beta'\beta}_{bg}(E)
 - i2\pi\delta^{4}(P_f-P_i)\sum_{\alpha,\alpha'} 
\phi^{\beta'\alpha'}(k;E)\Delta_{\alpha'\alpha}(E)^{-1}{\phi}^{\alpha\beta}(k;E)
\end{equation}
\end{widetext}
where $k$ is the on-shell momentum of the two meson state and the propagator is  
\begin{equation}
\Delta^{\alpha'\alpha}(E)=\left\{(E-M_\alpha)\delta^{\alpha'\alpha}+\mathcal 
G^{\alpha'\alpha}(E)\right\}
\end{equation}
Expanding around the pole, which now is defined as $|\Delta(\bar E)|=0$:
\begin{equation}
\begin{split}
 \Delta^{\alpha'\alpha}(E)-\Delta^{\alpha'\alpha}(\bar E)&=(E-\bar 
E)\left[\delta^{\alpha'\alpha}+\mathcal {G'}^{\alpha'\alpha}(\bar E)\right] \\ 
&=(E-\bar E)\mathcal Z^{\alpha'\alpha}(\bar E)
\end{split}
\end{equation}
with 
\begin{equation}
\mathcal {G'}^{\alpha'\alpha}(\bar E)=\lim_{E\to\bar E}\dfrac{\mathcal 
G^{\alpha'\alpha}(E)-\mathcal G^{\alpha'\alpha}(\bar E)}{E-\bar E} 
\end{equation}
Then the $S$-matrix can be approximated in the neighborhood of the pole as
\begin{widetext}
\begin{equation}
S^{\beta'\beta}(E)=
S^{\beta'\beta}_{bg}(E)
- i2\pi\delta^{4}(P_f-P_i) \sum_{\alpha,\alpha'} 
\phi^{\beta'\alpha'}(\bar k;\bar E)\dfrac{\mathcal Z_{\alpha'\alpha}(\bar E)^{-1}}{E-\bar E}
{\phi}^{\alpha\beta}(\bar k;\bar E)
\end{equation}
\end{widetext}
assuming that 
\begin{equation}
\mathcal Z_{\alpha'\alpha}(\bar E)=\sum_{\lambda}\mathcal 
Z_{\alpha'\lambda}^{1/2}\mathcal Z_{\lambda\alpha}^{1/2}
\end{equation}
the $S$-matrix can be finally written as
\begin{widetext}
\begin{equation}
S^{\beta'\beta}(E)=
S^{\beta'\beta}_{bg}(E)
- i2\pi\delta^{4}(P_f-P_i) \sum_{\alpha,\alpha',\lambda} 
\left[\phi^{\beta'\alpha'}(\bar k;\bar E)\mathcal 
Z_{\alpha'\lambda}(E)^{-1/2}\right]\dfrac{1}{E-\bar E}\left[\mathcal 
Z_{\lambda\alpha}(E)^{-1/2}{\phi}^{\alpha\beta}(\bar k;\bar E)\right].
\end{equation}
\end{widetext}
The vertex we are interested in is
\begin{equation}
S(X_c\rightarrow f)^{\beta\alpha}= \sum_{\lambda}\phi^{\beta\lambda}(\bar 
k;\bar E)\mathcal Z_{\lambda\alpha}(\bar E)^{-1/2}
\end{equation}
and the partial width can be defined as
\begin{equation} 
\label{ec:parcial1}
\hat \Gamma_f=\int d\Phi_f |S(X_c\rightarrow f)|^2
\end{equation}
where the integral is over the phase space of the final state with 
$\left(\sum_n p_n\right)^2=M_r^2$.

In the case of a two meson decay $\hat \Gamma_\beta$ can be written as
\begin{equation}
\label{ec:parcial2}
\begin{split}
 \hat\Gamma_\beta=&2\pi \dfrac{E_1E_2}{M_r}{k_0}_\beta \\ 
&\sum_{\alpha',\alpha,\lambda}{\phi^*}^{\beta\alpha'}(\bar k)\mathcal 
Z^*_{\alpha'\lambda}(\bar E)^{-1/2} \mathcal Z(\bar E)^{-1/2}_{\lambda 
\alpha}\phi^{\alpha\beta}(\bar k)
\end{split}
\end{equation}
where ${k_0}_\beta$ is the onshell momentum of the two meson state.

Eq.~(\ref{ec:parcial2}) does not guarantee that the sum of the partial 
widths must be equal to the total width. In fact it is expected that $\sum_f 
\hat \Gamma_f\neq \Gamma_r$. To solve this problem we define the 
$\emph{branching ratios}$ by~\cite{Grassi00}
\begin{equation} \label{ec:bratios}
\mathcal{B}_f=\frac{\hat\Gamma_f}{\sum_f\hat \Gamma_f}
\end{equation}
and the partial widths by
\begin{equation} 
  \Gamma_f=\mathcal{B}_f \Gamma_r.
\end{equation}

\section{Calculations, results and discussions}
\label{results}

Using the formalism developed in section~\ref{model} we have performed a coupled channel 
calculation of molecular states and $q\bar q$ pairs in different sectors of the 
charmonium spectrum. We use the parametrization of the $q\bar q$ interaction, 
together with the values for quark masses and the strength of the $^3P_0$ model 
of Ref.~\cite{segovia1}.

We have analyzed the positive parity sectors from 3.8 to 4.0 GeV where more of the new 
charmonium-like states appear. We have also calculate the effects of molecular 
structures for the $J^{PC}=1^{--}$ states in the region around 4.0 GeV.  

\subsection{Positive parity sector}

In Table~\ref{t1} we summarized the $XYZ$ candidates below 4.0 GeV that we will 
discuss in this work together with the most likely $J^{PC}$ assignments and its 
dominant decay modes. In Table~\ref{t2} we show the prediction in this region 
from the model of Ref.~\cite{segovia1}.

\begin{table}[!t]
\begin{center}
\begin{tabular}{ccccc}
\hline
\hline
State & M (MeV) & $\Gamma$ (MeV) & $J^{PC}$ & Decay mode \\
\hline
$X(3872)$ & $3871.4\pm0.6$ & $<2.3$ & $1^{++}$ & $D{D}^{\ast}$ \\
$X(3915)$ & $3914\pm3\pm2$ & $23\pm9$ & ? & $D^{\ast}{D}^{\ast}$ \\
$Z(3930)$ & $3929\pm5$ & $29\pm10$ & $2^{++}$ & $D{D}$ \\
$X(3940)$ & $3942\pm9$ & $37\pm17$ & $1^{++}$ & $D{D}^{\ast}$ \\
$Y(3940)$ & $3943\pm17$ & $87\pm34$ & ? $C=+$ & $J/\psi\omega$ \\
%$Y(4140)$ & $4143\pm$ & $11.7^{+8.0}_{-5.0}\pm3.7$ & ? $C=+$ & $J/\psi\phi$ \\
%$Y(4160)$ & $4156\pm29$ & $139^{+113}_{-65}$ & ? $C=+$ & 
%$D^{\ast}{D}^{\ast}$ (non $D{D}$) \\
\hline
\hline
\end{tabular}
\caption{\label{t1} Summary of candidates $XYZ$ mesons discussed in this work.}
\end{center}
\end{table}

\begin{table}[!t]
\begin{center}
\begin{tabular}{ccc}
\hline
\hline
State & M $(MeV)$ & $J^{PC}$ \\
\hline
%$\eta_{c}(3S)$ & $4054$ & $0^{-+}$ \\
$\chi_{c0}(2P)$ & $3909$ & $0^{++}$ \\
$h_{c}(2P)$ & $3955$ & $1^{+-}$ \\
$\chi_{c1}(2P)$ & $3947$ & $1^{++}$ \\
%$\eta_{c2}(2D)$ & $4166$ & $2^{-+}$ \\
$\chi_{c2}(2P)$ & $3968$ & $2^{++}$ \\
%$\chi_{c2}(1F)$ & $4045$ & $2^{++}$ \\
%$\psi_{c1}{2D}$ & $4152$ & $1^{--}$ \\
%$\psi_{c2}(2D)$ & $4164$ & $2^{--}$ \\
\hline
\hline
\end{tabular}
\caption{\label{t2} Prediction from our $c\bar c$ model in the region around $3970$ MeV.}
\end{center}
\end{table}

Although the model predicts four states in this region, the $h_c(2P)$ has 
negative $C$-parity and does not match with the data we are looking for. Then we 
have only three states predicted by the quark model whereas experimentally one 
found 5 states. 

As far as masses are concern, there is one clear identification: the 
$\chi_{c2}$ match the $Z(3930)$ mass (see Ref.~\cite{Seg01} for more 
properties). For the rest of states one can briefly comment that the 
$\chi_{c1}$ is too high in mass to be $X(3872)$. On the other hand the 
$X(3940)$ cannot be a $J^{P}=0^{++}$ state because, while a clear signal for 
$X(3940)\to D{D}^{\ast}$ is seen, there is no evidence for the $X(3940)$ in 
either $D{D}$ or $\omega J/\psi$ decay channels. Then the most likely 
candidate for $X(3940)$ is $J^{PC}=1^{++}$ with $M=3947$ MeV.
Concerning the $X(3915)$, although it has not been seen in the $D D$ channel due 
basically to its small branching ratio for this channel, it is a good candidate 
for our $\chi_{c0}$ state.

With this preliminary assignments our $c\bar c$ model predicts no candidates 
for the $X(3872)$ and the $Y(3940)$ if it finally exists.

The existence of two  $J^{PC}=1^{++}$ almost degenerated in mass, namely the 
$X(3940)$ and the $X(3872)$ suggest that the $1^{++}$ $c\bar{c}$ sector at 
these energies is more complicated than a simple $c\bar{c}$ structure. Moreover,
its extremely low binding energy makes the $X(3872)$ an ideal candidate to a 
$DD^*$ molecule.

In a earlier publication~\cite{ortega08} we performed a coupled channel 
calculation including the $c\bar c (2^3P_1)$ pair together with the neutral and 
charged $DD^*$ channels. In this work we found a bound state with an important 
molecular component. However one can wonder if the same effects can be found 
coupling the $\rho J/\psi$ and $\omega J/\psi$ neglected in this previous 
calculation.

To elucidate this point we have performed a full calculation including the two 
quark $c\bar c (2^3P_1)$ together with the $D^0D^{*0}$. $D^{\pm}D^{*\mp}$, 
$\rho J/\psi$ and $\omega J/\psi$ channels. The coupling of the $DD^*$ with the 
$\rho J/\psi$ and $\omega J/\psi$ channels is not enough to bind the molecule 
and it is mandatory to couple the $c\bar c$ pair to reach a molecular bound 
state.

One striking feature of the $X(3872)$ decays is the value of the ratio between the 
$X(3872)\rightarrow \rho J/\psi$ and the
$X(3872)\rightarrow \omega J/\psi$ decay channels. As stated above this ratio 
suggests that some isospin mixing is needed to reproduce the experimental data. 
To introduce the isospin breaking in our calculation we will work in the charge 
basis instead of in the isospin symmetric one, allowing the 
dynamics of the system to choose the weight of the different components. 
Isospin is explicitly broken by the experimental meson masses.

To have an idea of the sensitivity of the $X(3872)$ structure with the binding 
energy and having in mind that this ranges from $0.6$ MeV to $0.25$ MeV, we have 
fine-tuned the $^3P_0$ gamma parameter to get exactly this two energies. We can 
see the results in Table~\ref{t3}.

\begin{table} [!t]
\begin{center}
\begin{tabular}{ccccccc}
\hline
\hline
 $\gamma$ & $E_{bind}$ & $c\bar c(2 ^3P_1)$ & $D^0{D^*}^0$ & $D^
\pm{D^*}^\mp$ & 
$J/\psi\rho$ & $J/\psi\omega$ \\
\hline
   $0.231$   & $-0.60$ & $12.40$  & $39.24$  & $7.46$ & $0.49$ & $0.40$ \\
   $0.226$   & $-0.25$ & $8.00$  & $86.61$  & $4.58$ & $0.53$ & $0.29$ \\ 
\hline
\hline
\end{tabular}
\caption{\label{t3} Binding energy (in MeV) and
channel probabilities  (in $\%$)
for the $X(3872)$ states for two different values of the $\gamma$ parameter in 
the $^3P_0$ model.}
\end{center}
\end{table}

From this table one can see that the $X(3872)$ is predominately a $D^0{D^*}^0$ 
molecule with a small admixture (less than $10\%$) of the $c\bar c(2 ^3P_1)$ 
state  and the charged $D^\pm{D^*}^\mp$ component. The two channels $\rho J/\psi$ 
and $\omega J/\psi$, although important for the decays, are not 
significant with respect to the binding energy. 

The influence of the different components can be determine from the $X(3872)$ 
decays. The ratio
\begin{equation}
R_2=\frac{\Gamma(X(3872)\to\gamma\Psi(2S))}{\Gamma(X(3872)\to\gamma J/\psi)} 
\end{equation}
is sensitive to the $c\bar c(2 ^3P_1)$ component because in the molecular 
picture the radiative decay $\Gamma(X(3872)\to\gamma\Psi(2S))$ is suppressed~\cite{XL3}. 
The first evidence of this decay was reported by BaBar with a 
branching fraction 
$\frac{\Gamma(X(3872)\to\gamma\Psi(2S))}{\Gamma(X(3872)\to\gamma 
J/\psi)}=3.4\pm 1.4$~\cite{XL4} which suggested a rather large value of the 
$c\bar c(2 ^3P_1)$ component. However, in 2010, using a larger sample of the
$B\rightarrow X(3872)K$ decay, the Belle Collaboration has not found evidences 
for the radiative decay $\Gamma(X(3872)\to\gamma\Psi(2S))$ giving the upper 
limit $\mathcal{B.R.} < 2.1$ at $90\%$ C.L.~\cite{XL5}. In Table~\ref{t4} we 
show the width and the ratio for this two decays using the standard expression 
for the electric dipole transition~\cite{nora}.
One can see that the value of our $c\bar c$ component is small enough to 
accommodate at the experimental results.

\begin{table}[!t]
\begin{center}
\begin{tabular}{cccc}
\hline
\hline
 $E_{bind}$ & $\Gamma_{\gamma J/\psi}$ & $\Gamma_{\gamma\psi(2S)}$ & $R_2$ \\
\hline
    $-0.60$ & $8.15$  & $9,80$  & $1.20$ \\
    $-0.25$ & $5.25$  & $6.31$  & $1.20$ \\ 
\hline
\hline
\end{tabular}
\caption{\label{t4} Decay widths (in keV) of the 
the $X(3872)$ states and its ratio for two different values of the $\gamma$ 
parameter in the $^3P_0$ model.}
\end{center}
\end{table}

In Table~\ref{t5} we show the calculated width for the decays $X(3872)\to \pi^+ 
\pi^- J/\psi$ and $X(3872)\to \pi^+ \pi^-\pi^0 J/\psi$. The result for the ratio is not far 
from the experimental value $R_1=0.8\pm 0.3$~\cite{m9}. Although the absolute 
value of both decay widths varies with the $X(3872)$ binding energy their ratio is 
less sensitive.
It is worth to notice that to obtain this ratio 
(close to the experimental value) it is enough to have only less than  $30\%$ 
of $I=1$ component because, as explained in the introduction, the different phase 
spaces of the $\rho J/\psi$ and the $\omega J/\psi$ channels conspires with the 
$DD^*$ charged components to reproduce the experimental result.

\begin{table}[!t]
\begin{center}
\begin{tabular}{cccc}
\hline
\hline
 $E_{bind}$ & $\Gamma_{\pi^+ \pi^- J/\psi}$ & $\Gamma_{\pi^+ 
\pi^-\pi^0 J/\psi}$ & $R_1$ \\
\hline
    $-0.60$ & $27.61$  & $14.40$  & $0.52$ \\
    $-0.25$ & $24.18$  & $10.64$  & $0.44$ \\ 
\hline
\hline
\end{tabular}
\caption{\label{t5} Strong decay widths (in keV) of the 
the $X(3872)$ states and its ratio for two different values of the $\gamma$ 
parameter in the $^3P_0$ model.}
\end{center}
\end{table}

As stated above the formalism developed in Section~\ref{model} allows us to treat 
simultaneously bound states and resonances above and below the thresholds. When 
we look for resonances above the $DD^*$ threshold, we found a resonance  at 
$M=3941.8$ MeV and $\Gamma=89.9$ MeV, which can be identified with the $X(3940)$. The 
different components for this resonance together with the $X(3872)$ for the 
value $\gamma=0.226$ are shown in Table~\ref{t6}.

\begin{table}[!t]
\begin{center}
\begin{tabular}{cccccc}
\hline
\hline
 $Mass$ & $c\bar c(2 ^3P_1)$ & $D^0{D^*}^0$ & $D^
\pm{D^*}^\mp$ & 
$J/\psi\rho$ & $J/\psi\omega$ \\
\hline
   $3871.5$ & $8.00$  & $86.61$  & $4.58$ & $0.53$ & $0.29$ \\
   $3941.8$ & $61.09$  & $18.53$  & $16.85$ & $0.01$ & $3.52$ \\ 
\hline
\hline
\end{tabular}
\caption{\label{t6} Masses (in MeV) and
channel probabilities  (in $\%$)
for the $X(3872)$ and $X(3940)$ states.}
\end{center}
\end{table}

As seen from Table~\ref{t6} the $X(3940)$ has almost the same value of the neutral 
and charged $DD^*$ components. Therefore the isospin breaking has almost 
completely disappear and the resonance is basically $I=0$. This justifies that 
the coupling with the $\rho J/\psi$ channel is almost negligible. 

This resonance decay basically through the $DD^*$ component being the branching 
ratio for the different channels $\mathcal{B.R.}(X(3872)\to DD^*)=0.89$, 
$\mathcal{B.R.}(X(3872)\to \omega J/\psi)=0.1$ and $\mathcal{B.R.}(X(3872)\to 
\rho J/\psi)=3\cdot10^{-4}$.
In this way the puzzle between this two states with the same quantum numbers 
seems to be solved in a satisfactory way.
 
The situation in the $J^{PC}=0^{++}$ is also puzzling. Before the discovery of 
the $X(3915)$ signal, it was supposed that the $Y(3940)$ could be the $J^{PC}=1^{++}$ or 
$J^{PC}=0^{++}$ state. If we look to the Table~\ref{t2}, our $c\bar c$ model predicts 
a $J^{PC}=0^{++}$ state at 3909 MeV and so
the $Y(3940)$ seems to be too high. Despite the fact that one of the states is narrow 
(the $X(3915)$) and the other is broad (the $Y(3940)$), lately there has been 
a tendency to consider that both are the same state. 
We have performed the same calculation as before but for the $J^{PC}=0^{++}$ 
sector. We include the $c\bar c(2^3P_0)$ channel together with the molecular 
channels $DD$ (3736.05 MeV), $\omega J/\psi$ (3879.56 MeV), $D_s 
D_s$ (3936.97 MeV), $\phi J/\psi$ (4116.0 MeV) where the channel thresholds are 
given between parenthesis.  
The results are shown in Table~\ref{t7}.

We find a narrow state which can be identified with the $X(3915)$. This 
resonance is basically a mixture of $c\bar c$ and a $DD$ molecular 
component. Moreover a second wide resonance appears at $M=3970$ MeV. This resonance 
can be the old $Y(3940)$, today disappeared from the Particle Data Group. Its dominant component is 
the $c\bar c(2 ^3P_0)$ although the contribution of the molecular  $DD$ 
one is also significant. This structure provide a possible explanation for the 
unusual decay mode $\omega J/\psi$ through the rescattering $0^{++}\rightarrow 
DD\rightarrow \omega J/\psi$.

\begin{table}[!t]
\begin{center}
\begin{tabular}{cccccccc}
\hline
\hline
 $ state$ & $Mass$ & $\Gamma$ &  $c\bar c(2^3P_0)$ & $D^0 \bar D^0$ & 
$J/\psi\omega$ & $D_s\bar D_s$ & $J/\psi\phi$ \\
\hline
   $X(3915)$   & $3896.5$ & $4.10$ & $34.22$  & $46.67$  & $9.42$ & $9.67$ & 
$0.03$ \\ 
   $Y(3940)$   & $3970$ & $189.3$ & $57.27$  & $35.32$  & $0.15$ & $5.72$ & 
$1.54$ \\ 
\hline
\hline
\end{tabular}
\caption{\label{t7} Mass and total width (in MeV) and
channel probabilities  (in $\%$)
for the $X(3915)$ and $Y(3949)$.}
\end{center}
\end{table}

Then our calculation shows that the $X(3915)$ and the $Y(3940)$ may be two 
different resonances as measured by Belle.

\subsection{Negative parity sector}

In the $J^{PC}=1^{--}$ sector one can find a similar situation to those studied 
in the previous sector. In fact the $\psi(4040)$ resonance with a mass just 
above the $D^*D^*$ threshold has been proposed long ago~\cite{XL6} as a 
candidate to a molecular state. To asses this possibility we have performed a 
coupled channel calculation including the $c\bar c(3^3S_1)$ and $c\bar c(2^3D_1)$  
states with masses $M=4097.615$ MeV and $M=4152.715$ MeV respectively, 
together with the channels $DD$, $DD^*$, $D^*D^*$, $D_sD_s$, 
$D_s D_{s}^*$ and $D_s^* D_s^*$. The results of the calculation are shown in 
Table~\ref{t8}. 

The first striking outcome of the calculation (see Table~\ref{t8}) is the 
appearance of a narrow state at $M=3994.6$ MeV. A state with this 
characteristics has been reported by BaBar~\cite{XL7} in the study of 
exclusive initial-state-radiation production of the $DD$ system. Its
experimental mass and width are respectively $M=3943\pm 17\pm 12$ MeV and 
$\Gamma=52\pm 8\pm 7$ MeV. This resonance has also been seen by the Belle 
Collaboration~\cite{XL8}.
The second significant result is that, due to the molecular mixing, the $c\bar 
c(2^3D_1)$ state becomes the most important component of the $\psi(4040)$  and 
not the $c\bar c(3^3S_1)$ as usually attributed by the naive quark model. 
Moreover the $\psi(4040)$ acquires a significant $D D^*$  ($23.49\%$) and  
$D^*D^*$ ($25.81\%$) components  while the  
$\psi(4160)$ is predominantly a $c\bar c(3^3S_1)$ state with small 
contributions ($< 20\%$) of different $DD$ and $D_sD_s$ channels.

\begin{table*}[!t]
\begin{center}
\begin{tabular}{ccccccccc}
\hline
\hline
 $ M(MeV)$ & $c\bar c 3^3S_1$ &  $c\bar c 2^3D_1$ &  $D \bar D$ & $D \bar D^*$ 
& $D^* \bar D^*$ & $D_s\bar D_s$ & $D_s\bar D_s^*$  & $D_s^*\bar D_s^*$ \\
\hline
   $3994.6-i11.60$   & $31.56$ & $3.0$ & $2.49$  & $36.44$  & $117.75$ & $7.53$ 
& $0.523$ & $0.71$ \\ 
   $4048.4-i7.54$   & $0.92$ & $36.15$ & $2.99$  & $23.49$  & $25.81$ & $8.86$ 
& $0.924$ & $0.85$  \\
   $4123.9-i71.11$  & $59.01$ & $0.98$ & $2.13$  & $6.84$  & $19.119$ & $0.75$ 
& $3.37$ & $7.73$  \\
\hline
\hline
\end{tabular}
\caption{\label{t8} Mass (in MeV) and
channel probabilities  (in $\%$)
for the $J^{PC}=1^{--}$ sector in the 4.0 GeV region.}
\end{center}
\end{table*}

These new assignments have an important influence on the decay ratios of 
these two resonances. It is well known that the ratios of branching fractions 
involving these two resonances shows significant discrepancies with model 
predictions, specially with the $^3P_0$ model. The predictions of our 
calculation are shown in Table~\ref{t9} together with two different quark 
models. As seen, the new assignment improve the overall agreement with the 
experimental data although some discrepancies remains.

\begin{table*}[!t]
\begin{center}
\begin{tabular}{ccccc}
\hline
\hline
 $ratio$ & $Measurements$ &$ ^3P_0$ \cite{XL9}& $C^3$ \cite{XL10} & This work \\
\hline
 $\mathcal{B.R.}(\psi(4040)\rightarrow D\bar 
D$)/$\mathcal{B.R.}(\psi(4040)\rightarrow D\bar D^*$)
  & $0.24\pm 0.05\pm 0.12$ & $0.003$  & $0.0003$  & $0.07$ \\
 $\mathcal{B.R.}(\psi(4040)\rightarrow D^*\bar 
D^*$)/$\mathcal{B.R.}(\psi(4040)\rightarrow D\bar D^*$)
 &   $0.18\pm 0.14\pm 0.03$ & $1.0$  & $1.0$  & $0.61$ \\ 
 $\mathcal{B.R.}(\psi(4160)\rightarrow D\bar 
D$)/$\mathcal{B.R.}(\psi(4160)\rightarrow D^*\bar D^*$)
 &   $0.02\pm 0.03\pm 0.02$ & $0.46$  & $0.008$  & $0.0.5$ \\
 $\mathcal{B.R.}(\psi(4160)\rightarrow D\bar 
D^*$)/$\mathcal{B.R.}(\psi(4160)\rightarrow D^*\bar D^*$)
 &   $0.34\pm 0.14\pm 0.05$ & $0.011$  & $0.16$  & $0.08$ \\
\hline
\hline
\end{tabular}
\caption{\label{t9} Ratios of branching fractions for the two $\psi$ 
resonances. Experimental data are from ref \cite{XL11}.}
\end{center}
\end{table*}

Another consequence of the important molecular component of the 
$\psi(4040)$ is that it can give rise to an enhancement of some specific decay 
channels like the $\psi(4040)\rightarrow \eta J/\psi$ through the presence of 
$(D^* D^*)_{S=0}$ pairs in the $\psi(4040)$ internal structure. This 
mechanism has been recently proposed by Voloshin~\cite{XL12} as a tool to 
identified possible molecular structures in this states.

\section{Summary}
\label{summary}

During the years charmonium spectrum below $DD$ threshold was a well 
described system in quark models. However since the last 10 years new states 
above this threshold has been measured with properties which can be hardly 
described in naive quark models. Obviously at these energies threshold effects 
has to be taken into account which sometimes has been referred as an unquenching 
of the quark model.

In this paper we address this issue focusing on the possible influence of the 
molecular structures on the charmonium spectrum. In the framework of a 
constituent quark model, we have developed a formalism which allows us to coupled 
$c\bar c$ states with four quark molecular states below and above the different 
thresholds. We study the positive parity sector in the mass region of the 
$X(3872)$ and the negative parity sector around masses of 4.0 GeV.

We describe the $X(3872)$ resonance as a $J^{PC}=1^{++}$ mixture of neutral and 
charged $DD^*$ molecular states and a less than $10\%$ contribution of the 
$c\bar c(2^3P_1)$ state what however  is enough to describe the electromagnetic 
decays of the resonance. The isospin breaking showed by the data is also well 
explained with this configuration. Together with this resonance in the 
$1^{++}$ sector appears a second state whose properties are compatible with 
those of the $X(3940)$ state.

We found two resonances with $J^{PC}=0^{++}$ quantum numbers. The first one, 
with important $c\bar c$ and $D D$ components, may be identified with the $X(3915)$. 
We found also a second broad resonance in the mass region $M=3940$ MeV which 
could be assign to the signal seen by the Belle Collaboration and which for 
sometime was called $Y(3940)$, although recently, probably due to the 
lack of quantum numbers to accommodate this resonance, was subsummed  under the 
name $X(3915)$.

Concerning the negative parity sector, we confirm the old suggestion of De 
R\'ujula {\it et al.}~\cite{XL6} that the $\psi(4040)$ resonance is mostly a molecular 
state. Moreover the coupling with these molecular structures change the $c\bar c$.
quantum numbers of the $\psi(4160)$ which acquires an important $^3S_1$ 
component, contrary to the usual hypothesis that is a $^3D_1$ state. This new 
assignment has important consequences on the decay branching ratios.
Finally the molecular components of these two resonances open the possibility 
of enhance the probability of new decay channels to detectable levels which 
deserves further studies.

\begin{acknowledgments}

This work has been partially funded by Ministerio de Ciencia y Tecnolog\'ia
under Contract Nos.  FPA2010-21750-C02-02, by the
European Community-Research Infrastructure Integrating Activity 'Study of
Strongly Interacting Matter' (HadronPhysics2 Grant No. 227431) and by the
Spanish Ingenio-Consolider 2010 Program CPAN (CSD2007-00042). 

\end{acknowledgments}

\end{document}